\documentclass[a4paper, aps, prd,nofootinbib, notitlepage, 10pt]{revtex4-1}
\usepackage[OT1,T1]{fontenc}
\usepackage[latin1]{inputenc}
\usepackage{amsmath}
\usepackage{amsfonts}
\usepackage{mathtools}
\usepackage{bbm}
\usepackage{accents}
\usepackage{dsfont}
\usepackage{mathrsfs}
\usepackage[dvips]{graphicx}
\usepackage{color}
\usepackage{graphicx}
\usepackage[normalem]{ulem}
\newcommand{\qq}[1]{``#1''} 
\usepackage{hyperref}
\newenvironment{subalign}{\subequations\align}{\endalign\endsubequations}
\newcommand{\utilde}[1]{\undertilde{#1}}
\newcommand{\di}{\mathrm{d}} 
\newcommand{\ou}[3]{{#1}{}^{#2}{}_{#3}} 
\newcommand{\uo}[3]{{#1}{}_{#2}{}^{#3}} 
\newcommand{\I}{\mathrm{i}} 
\newcommand{\E}{\mathrm{e}} 
\newcommand{\ellp}{{\ell_{\mathrm{P}}}} 
\newcommand{\CC}{\mathrm{cc.}} 
\newcommand{\lbreck}{[}
\newcommand{\rbreck}{]}
\newcommand{\eref}[1]{(\ref{#1})}

\newcommand{\C}{\mathbb{C}}
\newcommand{\N}{\mathbb{N}}
\newcommand{\R}{\mathbb{R}}
\newcommand{\Z}{\mathbb{Z}}

\newcommand{\bra}[1]{\langle{#1}|}
\newcommand{\ket}[1]{|{#1}\rangle}
\newcommand{\robra}[1]{\lbreck{#1}|}
\newcommand{\roket}[1]{|{#1}\rbreck}

\begin{document}
\title{A one-dimensional action for simplicial gravity in three dimensions}
\author{Wolfgang M. Wieland}
\email{wieland@gravity.psu.edu}
\affiliation{
Institute for Gravity and the Cosmos \& Physics Department\\
104 Davey Lab, PMB \#070\\
University Park, PA 16802, U. S. A.
}

\date{February 2014}
\begin{abstract}
\noindent This article presents a derivation of the Ponzano--Regge model from a one-dimensional spinor action. The construction starts from the first-order Palatini formalism in three dimensions. We then introduce a simplicial decomposition of the three-dimensional manifold and study the discretised action in the spinorial representation of loop gravity. A one-dimensional refinement limit along the edges of the discretisation brings us back to a continuum formulation. The three-dimensional action turns into a line integral over the one-skeleton of the simplicial manifold. All fields are continuous but have support only along the one-dimensional edges. We define the path integral, and remove the redundant integrals over the local gauge orbits through the usual Faddeev--Popov procedure. The resulting state sum model reproduces the Ponzano--Regge amplitudes.
\end{abstract}
\maketitle
\section{Introduction}\noindent
Three-dimensional gravity is topological, there are no propagating degrees of freedom, and yet it is rich enough to make its quantisation an intriguing problem \cite{carlipbook,Witten198846,ponzanoregge,Deser1984220}. 
Solving this problem is an important consistency check for any approach that aims at quantum gravity in the real world.

This article provides such a consistency check for the spinorial representation of loop gravity, recently developed by Freidel, Speziale and collaborators \cite{twist,Livinerep,spezialetwist1,Borja:2010rc,twistcons,twistintegrals,Dupuis:2012vp,komplexspinors, hamspinfoam,Mingyitwist,Smolin:2013mma}. 
The spinorial framework sits half way in between the most familiar connection representation \cite{LOSTtheorem, status, thiemann, rovelli}, and the dual Baratin--Oriti momentum representation \cite{Baratin:2010nn, Dittrich:2014wpa}. 
The spinors start from a different polarisation of the phase space of the theory, and parametrise at the same time both holonomies and fluxes. 
Spinors simplify the kinematical structure of the theory, but can they also teach us something about the dynamics? Here, we study this question only for the case of Euclidean gravity in three dimensions, and find a neat derivation of the Ponzano--Regge model \cite{ponzanoregge} from a one-dimensional spinor action. 

The article develops two results. Section \ref{newactn} gives the classical part. We discretise the first-order Palatini action $S_M$ over a simplicial decomposition of the underlying manifold. A one-dimensional refinement limit brings us back to a continuum formulation. The resulting action is a line integral over the edges of the simplicial decomposition (figure \ref{tetra}). We can rearrange this action so as to get a sum over the elementary spinfoam faces $f$, each of which contributes as follows:
\begin{equation}\nonumber
S_M=\sum_{f:\text{faces}}S_f,\quad\text{with:}\quad S_f=-\I\hbar\int_{\partial f}\Big(\langle z|D|z\rangle - \langle w|\di|w\rangle-\I\,\varphi\,\di t\,\big(\langle z|z\rangle-\langle w|w\rangle\big)\Big).\label{spinactn}
\end{equation}
The action $S_M$ depends on two spinors for each face, but is also a functional of an $SU(2)$ connection $A$ hiding in the covariant differential $D|z\rangle=\di|z\rangle+A|z\rangle$ of the spinor (and $\varphi$ is a Lagrange multiplier, while $t$ parametrises the boundary of $f$). All fields are continuous, but are supported only along the one-dimensional edges of the discretisation. 
Next, we study the local gauge symmetries of the theory, and derive the equations of motion from the principle of least action. The resulting theory is a version of first order Regge calculus, with spinors as the fundamental configuration variables.

Section \ref{pathintsec} develops the second result, and defines the transition amplitudes as a path integral over the spinorial variables: 
\begin{equation}\nonumber
Z_{\text{PR}}=\int\mathcal{D}_{\text{gf}}[z,w,\dots]\prod_{f:\text{faces}}\E^{\frac{\I}{\hbar}S_f}.
\end{equation}
The integration measure includes a gauge fixing condition together with the corresponding Faddeev--Popov determinant. This removes the divergent integrals over the orbits of the local gauge symmetries. We evaluate the integral for a generic simplicial decomposition and establish the equivalence with the Ponzano--Regge spinfoam model. This is our final result. It proves that we can derive the Ponzano--Regge spinfoam model from a one-dimensional spinorial field theory over the one-skeleton\footnote{More precisely: The one-skeleton of the dual complex. This is the system of edges glued among the bounding vertices. See figure \ref{tetra} for an illustration.} of the simplicial manifold.

\section{Euclidean gravity in three dimensions}
\noindent The entire section is a review, needed to make the article logically self-contained. Section \ref{3dgravi} introduces the most basic mathematical structures underlying three-dimensional Euclidean gravity.  Section \ref{loopvariables} gives the phase space for the discretised theory \cite{EEcomm}. The concluding section \ref{spinrep} studies the spinorial representation of loop gravity as developed by Freidel, Speziale and collaborators \cite{twist,Livinerep,spezialetwist1,Borja:2010rc,twistcons,twistintegrals,Dupuis:2012vp,komplexspinors, hamspinfoam,Mingyitwist}. References \cite{carlipbook,ReikoWDW, Barrett:2008wh, ashtekar,alexreview} give further background material.
\subsection{First-order action and simplectic structure}\label{3dgravi}\noindent
We are using first-order variables. The action for Euclidean gravity on a three-dimensional manifold $M$ thus becomes:
\begin{equation}
S_M[e,A]=\frac{\hbar}{2\ellp}\int_M\epsilon_{ijk}e^i\wedge F^{jk},\label{actn}
\end{equation}
where $\ellp$ and $\hbar$ are the Planck length and Planck's constant respectively, the flat Euclidean metric $\delta_{ij}$ moves all internal $\R^3$-indices $i,j,k,\dots$, and $\epsilon_{ijk}$ is the Levi-Civita tensor in internal space. The action is a functional both of the $\mathfrak{so}(3)$ connection $\ou{A}{i}{j\mu}$ and the cotriad $\ou{e}{i}{\mu}$. The cotriad is an orthonormal frame, it diagonalises the Euclidean line element $g_{\mu\nu}=\delta_{ij}\ou{e}{i}{\mu}\ou{e}{j}{\nu}$.
The $\mathfrak{so}(3)$ connection $\ou{A}{i}{j}$ defines the curvature two-form $
\ou{F}{i}{j}=\di\ou{A}{i}{j}+\ou{A}{i}{k}\wedge\ou{A}{k}{j}
$. 
We can equally well work with an $\mathfrak{su}(2)$ connection instead. The isomorphism between $\mathfrak{so}(3)$ and $\mathfrak{su}(2)$ is given by $\ou{A}{i}{j}=\ou{\epsilon}{i}{kj}A^k\mapsto A=A^k\otimes\tau_k$, where $\tau_k$ is a basis in 
$\mathfrak{su}(2)$ such that $[\tau_i,\tau_j]=\uo{\epsilon}{ij}{k}\tau_k$. If $\sigma_i$ are the Pauli matrices, a possible choice is $\tau_i=(2\I)^{-1}\sigma_i$. The action variation gives the equations of motion, namely:
\begin{subequations}
\begin{align}
\text{the torsionless condition:}\quad\ou{T}{i}{\mu\nu}&=2D_{[\mu}\ou{e}{i}{\nu]}=0,\label{torscons}\\
\text{and the flatness constraint:}\quad\ou{F}{i}{\mu\nu}&=0,\label{flatcons}
\end{align}\label{conteom}
\end{subequations}
where $D=\di+[A,\cdot]$ is the exterior covariant derivative, and $[\mu\dots]$ denotes anti-symmetrisation of all intermediate indices.
The unique solution of the torsionless condition $De^i=0$ determines the $SU(2)$ connection as a functional of the triad: The $SU(2)$ connection $\ou{A}{i}{\mu}$ turns into the Levi-Civita spin connection $\ou{\Gamma}{i}{\mu}[e]$. The equation $\ou{F}{i}{\mu\nu}=0$, on the other hand, tells us that the curvature of the connection vanishes, hence the metric $g_{\mu\nu}=e_{i\mu}\ou{e}{i}{\nu}$ is locally flat.

We want to eventually quantise the theory, so let us briefly recapitulate those aspects of its Hamiltonian formulation that will become important for us. We start with a 2+1 split of the three-dimensional manifold $M$, and foliate $M\simeq\Sigma\times \R$ into $t=\mathrm{const.}$ equal \qq{time} slices $\Sigma_t\simeq\Sigma\times\{t\}$. The 2+1 decomposition requires a time-flow vector field\footnote{$\mu,\nu,\rho,\dots$ ($a,b,c,\dots$) are abstract indices in $TM$ ($T\Sigma$).} $t^\mu\in TM$, transversal to the $t=\mathrm{const}.$ hypersurfaces: $t^\mu\partial t=1$. Once we have chosen such a vector field, we can define the spatial and \qq{temporal} components of the configuration variables:
\begin{equation}
\begin{split}
N^i&:=t^\mu\ou{e}{i}{\mu},\\
\ou{e}{i}{a}&:=[\mathrm{em}_t^\ast e^i]_a,
\end{split}\qquad
\begin{split}
\phi^i&:=t^\mu\ou{A}{i}{\mu},\\
\ou{A}{i}{a}&:=[\mathrm{em}^\ast_t A^i]_a.
\end{split}
\end{equation}
We are working with Euclidean geometries, therefore this time function has no physical meaning whatsoever. 
Moreover, $\mathrm{em}_t:\Sigma\rightarrow\Sigma_t;x\mapsto (x,t)\in M$ is the canonical embedding of $\Sigma$ into $M$, and $\mathrm{em}_t^\ast$ is the corresponding pull-back: $\ou{e}{i}{a}$ and $\ou{A}{i}{a}$ are fields intrinsically defined on $\Sigma$, they are the pull-back of the three-dimensional fields $\ou{e}{i}{\mu}$ and $\ou{A}{i}{\mu}$ to the $t=\mathrm{const.}$ slice. We also define the velocity $\ou{\dot{A}}{i}{a}=[\mathrm{em}_t^\ast(\mathcal{L}_tA^i)]_a$ of the connection as the pull-back of the Lie derivative $\mathcal{L}_t\ou{A}{i}{\mu}$. If we now also introduce the covariant derivative $D_a$ with respect to $\ou{A}{i}{a}$ on $\Sigma$, and call $\ou{F}{i}{ab}=[\mathrm{em}_t^\ast F^i]_{ab}=[D_a,D_b]^i$ its curvature, then we can write down the action\footnote{To evaluate the integral we need to speak about orientation. Assume that $M$ be orientable, and so be $\Sigma_t$: If $(t,X,Y)$ are positively oriented vector fields in $M$, we choose the orientation in $\Sigma_t$ such that the duple $(X,Y)$ has positive orientation.} in the following canonical form:
\begin{equation}
S[e,A]=\frac{\hbar}{\ellp}\int_0^1\di t\int_\Sigma\tilde{\eta}^{ab}\Big(e_{ia}\big(\ou{\dot{A}}{i}{b}-D_b\phi^i\big)-\frac{1}{2}N_i\ou{F}{i}{ab}\Big).
\end{equation}
Here $\tilde{\eta}^{ab}$ is the Levi-Civita density; its inverse (a density of weight minus one) is $\utilde{\eta}_{ab}$ (and $\utilde{\eta}_{ac}\tilde{\eta}^{bc}=\delta^b_a)$. Looking at the first term in the action, we can identify the symplectic structure; the only non-vanishing Poisson brackets are:
\begin{equation}
\big\{\ou{e}{i}{a}(x),\ou{A}{j}{b}(y)\big\}=\frac{\ellp}{\hbar}\delta^{ij}\utilde{\eta}_{ab}\tilde{\delta}_\Sigma(x,y),\label{contpoiss}
\end{equation}
where $\tilde{\delta}_\Sigma(x,y)$ is the Dirac distribution on the two-dimensional $t=\mathrm{const}.$ slice $\Sigma$, a scalar density of weight one.

The canonical coordinates in the phase space of the theory are thus an $\mathfrak{su}(2)$-valued one-form $\ou{e}{i}{a}$ and an $SU(2)$ connection $\ou{A}{i}{a}$. These are the spatial projections of the configuration variables $e^i$ and $A^i$ in the action \eref{actn}. The temporal components $N^i$ and $\phi^i$ play a different role, they appear as Lagrange multipliers, and impose the constraints of the theory, which are nothing but the equations of motion \eref{conteom} pulled-back to the spatial slice. 
\subsection{Holonomy-flux variables}\label{loopvariables}
\noindent In loop gravity \cite{ashtekar, rovelli, thiemann, alexreview} we work on a truncated phase space of smeared variables. We can think of this truncation as the result of a discretisation: All fields are discretised over the elementary building blocks of a triangulation\footnote{In principal we do not have to stick to triangulations, the kinematics of loop gravity allows arbitrary polytopes \cite{polyhdr}.} of the three-dimensional manifold \cite{EEcomm, contphas}. 

\begin{figure}[h]
     \centering
     \includegraphics[width= 0.3\textwidth]{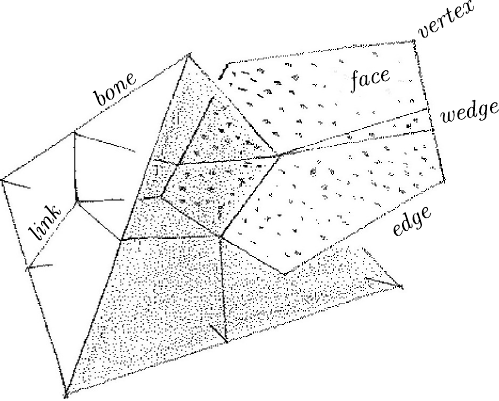}
     \caption{A tetrahedron consists of four triangles glued together. Each tetrahedron contains its own dual, the \emph{vertex}, a point inside. Three sides bound a triangle, we call them the \emph{bones} of the triangulation. Each bone belongs to many tetrahedra (vertices), but a triangle can only be in two of them. The surface dual to a bone is a \emph{face}, it touches all adjacent tetrahedra. An \emph{edge}, the dual of a triangle, connects two vertices. A wedge is a \qq{small} triangular part of a face: Two of its corners belong to an edge, the third lies on the bone dual to the face. From the two-dimensional perspective of e.g. the boundary of a tetrahedron, a \emph{link} is the dual of a bone. A wedge is thus bounded by two links and a short segment of an edge.}
     \label{tetra}
\end{figure}

We thus introduce a simplicial decomposition of $M$, which consists of \emph{tetrahedra} $T$ glued along their bounding \emph{triangles} $\tau\subset \partial T$ (see figure \ref{tetra} for an illustration). Each triangle bounds two tetrahedra, and is itself bounded by three sides, which we call the \emph{bones} $b\subset\partial \tau$ of the triangulation. It is also important to know about the dual picture: Each tetrahedron is dual to a \emph{vertex} (a point $v\in M$), while each triangle is dual to an \emph{edge} $e$ (a one-dimensional line). Edges close to form two-dimensional surfaces. These are the \emph{faces} $f$, each of which is dual to a bone. We can then use the three-dimensional discretisation of $M$ to triangulate a two-dimensional hypersurface $\Sigma\subset M$. This time, the elementary building blocks are just triangles glued along their bounding sides. From this two-dimensional perspective, every triangle is dual to a \emph{node} (a point in $\Sigma$), and every bone $b, b^\prime,\dots$ is dual to a \emph{link} $\gamma,\gamma^\prime,\dots$ (a path in $\Sigma$). At each node three links meet, and every link connects two adjacent nodes. 

The elementary building blocks of the triangulation are oriented: Each face $f$ carries an orientation, but the orientation of its bounding edges $e$ is independent, and may not match the induced orientation of $\partial f$. Furthermore, every bone has an orientation such that it is positively oriented relative\footnote{If $Z$ is a tangent vector in $b$, and $(X,Y)\in TM\times TM$ is a positively oriented duple in $f$, then the triple $(Z,X,Y)$ shall be positively oriented in $M$.} to its dual face $f$. There are also the oppositely oriented elements, we denote them $f^{-1}$, $b^{-1}$ and so on. Consider now a two-dimensional oriented hypersurface $\Sigma\subset M$ formed by glueing together adjacent triangles. We have already introduced the links $\gamma, \gamma^\prime,\dots\subset\Sigma$, each of which is dual (from the two-dimensional perspective of $\Sigma$) to a bone $b, b^\prime,\dots\subset \Sigma$, we now give them an orientation and demand that the duple $(\dot\gamma,\dot b)$ of corresponding tangent vectors be positively oriented in $\Sigma$.

Next, we introduce the smearing. We take the oriented bones $b, b^\prime,\dots$ and their dual in $\Sigma$ (the links $\gamma, \gamma^\prime,\dots$), and smear the elementary phase space variables $\ou{e}{i}{a}$ and $\ou{A}{i}{a}$ over these lower dimensional structures. The connection defines the parallel propagator between any two nodes as the path-ordered exponential:
\begin{equation}
\text{holonomy:}\quad h[b]=\mathrm{Pexp}\Big(-\int_\gamma A\Big)\in SU(2),\label{holdef}
\end{equation}
where $b$ is the bone dual to the link $\gamma$.
This gives the smearing of the connection. For the triad the situation is a little more complicated. The triad is a one-form, and we can smear it over the bones of the triangulation. The naive definition $\ell^i[b]=\int_b e^i$ breaks however $SU(2)$ gauge invariance, because it does not make sense to add internal vectors that belong to different points in $b\subset \Sigma$. The solution is to introduce additional holonomies $h_{\delta(x\rightarrow\gamma(0))}\in SU(2)$ that transport any internal vector in $x\in b$ into the frame at the initial point of $\gamma$:
\begin{equation}
\text{flux:}\quad\ell[b]=\int_b e^i(x)\, h_{\delta(x\rightarrow\gamma(0))}\tau_ih_{\delta(x\rightarrow\gamma(0))}^{-1}\in\mathfrak{su}(2).\label{fluxdef}
\end{equation}
The underlying path $\delta(x\rightarrow\gamma(0))$ starts at $x\in b$ follows the bone to the intersection point $b\cap \gamma$, where it then leaves $b$, and goes along $\gamma^{-1}$ until it reaches the source $\gamma(0)$.

Let us also mention the oppositely oriented elements. Changing the orientation amounts to replace the loop variables according to the following scheme:
\begin{equation}
h[b^{-1}]=h[b]^{-1},\quad\utilde{\ell}[b]:=\ell[b^{-1}]\equiv\ell^i[b^{-1}]\tau_i=-h[b]\,\ell[b]\,h[b]^{-1}.\label{orientdef}
\end{equation}

The commutation relations of the continuum theory \eref{contpoiss} induce commutation relations for holonomies and fluxes: 
\begin{subequations}
\begin{align}
\big\{\ou{h}{A}{B}[b],\ou{h}{C}{D}[b^\prime]\big\} & = 0,\\
\big\{\ell_i[b],\ou{h}{A}{B}[b^\prime]\big\} & = +\frac{\ellp}{\hbar}\delta_{bb^\prime}\ou{h}{A}{C}[b]\ou{\tau}{C}{Bi},\\
\big\{\ell_i[b],\ell_j[b^\prime]\big\} & = +\frac{\ellp}{\hbar}\delta_{bb^\prime}\uo{\epsilon}{ij}{m}\ell_m[b].
\end{align}\label{holfluxal}
\end{subequations}
Variables belonging to different links commute, and the algebra closes. The resulting phase space is nothing but the cotangent bundle $T^\ast SU(2)^L$ equipped with its natural symplectic structure ($L$ counts the number of links in the triangulation).

\subsection{Spinors for loop gravity}\label{spinrep}\noindent
Before we continue our review, and speak about loop gravity in the spinorial representation \cite{twist, Livinerep}, let us first fix some conventions. We will mostly use an index notation and denote the spinors as elements $z^A, w^A, \dots$ of $\C^2$ with $A\in\{0,1\}$ labelling their \qq{up} and \qq{down} components. There is also the complex conjugate vector space $\bar{\C}^2$, an overbar decorates the corresponding indices: $\bar{z}^{\bar A}\in\bar{\C}^2$. Spinors carry a natural action of $SU(2)$, the group acts through its fundamental matrix representation: $SU(2)\ni U:z^A\mapsto (Uz)^A=\ou{U}{A}{B}z^B$. Elements of $SU(2)$ are both unimodular and Hermitian, thus implying that both the anti-symmetric $\epsilon$-tensor and the Hermitian metric $\delta_{A\bar A}$ commute with the group action. We can thus invariantly move the spinor indices according to the following scheme:\footnote{We give the only non-vanishing components of the invariant tensors: $\delta_{0\bar 0}=1=\delta_{1\bar 1}$ and $\epsilon_{01}=1=-\epsilon_{10}$, the inverse of the $\epsilon$-tensor is defined implicitly: $\epsilon^{BC}\epsilon_{AC}=\epsilon_A{}^B=\delta^B_A.$} 
\begin{equation}
\begin{split}
\robra{z}&=z_A=\epsilon_{BA}z^B,\\
\roket{z}&=z_\dagger^A=\delta^{A\bar A}\bar{z}_{\bar A},
\end{split}\qquad
\begin{split}
\ket{z}&=z^A=\epsilon^{AB}z_B,\\
\bra{z}&=z^\dagger_A=\delta_{A\bar A}\bar{z}^{\bar A},
\end{split}\label{inmov}
\end{equation}and $\langle z |z\rangle=[z|z]=\|z\|^2=\delta_{A\bar A}z^A\bar{z}^{\bar A}$ denotes the corresponding $SU(2)$-norm. Notice also, that the intertwining maps \eref{inmov} generalise naturally to any higher rank spinor $T^{ABC\dots}$.

We now use these $SU(2)$ spinors to parametrise both holonomy and flux. The flux $\ell[b]$ is an element of $\mathfrak{su}(2)$, it defines an anti-Hermitian $2\times 2$ matrix $\ell[b]=\ell^A{}_B[b]=\ell^i[b]\ou{\tau}{A}{Bi}$, and thus has two orthogonal eigenspinors $\ket{z}=z^A$ and $\roket{z}=z_\dagger^A$. Their normalisation is free, and we can conveniently choose it to measure the metrical length of $b$ in units of the Planck length $\ellp$:
\begin{equation}
\ell[b]=+\frac{\ellp}{4\I}\Big(\ket{z}\bra{z}-\roket{z}\robra{z}\Big),\quad\ell_{AB}[b]=+\frac{\ellp}{2\I}z^{\phantom{\dagger}}_{(A}z^\dagger_{B)},\label{fluxpar}
\end{equation} 
where $(A\dots)$ denotes symmetrisation of all intermediate indices. Normalised like this, the spinors are unique up to an overall $U(1)$ transformation $z^A\mapsto\E^{\I\Omega}z^A$. They belong to the $SU(2)$ frame at the initial point. In the frame at the final point \eref{orientdef} we can find another pair of diagonalising spinors:
\begin{equation}
\utilde{\ell}[b]=-\frac{\ellp}{4\I}\Big(\ket{\utilde{z}}\bra{\utilde{z}}-\roket{\utilde{z}}\robra{\utilde{z}}\Big),\quad\utilde{\ell}_{AB}[b]=-\frac{\ellp}{2\I}\utilde{z}^{\phantom{\dagger}}_{(A}\utilde{z}^\dagger_{B)}.\label{fluxpar2}
\end{equation} 
The holonomy maps the flux $\ell[b]$ at the initial point into the flux $\utilde{\ell}[b]$ at the final point, equation \eref{orientdef} gives the precise relation. The spinors are unique up to an overall face, and therefore equation \eref{orientdef} translates into the following condition:
\begin{equation}
\exists\Phi\in\R:\utilde{z}^A=\E^{\I\Phi}\ou{h}{A}{B}[b]z^B.\label{spinrel}
\end{equation}
There is thus an $SU(2)$ transformation that maps one spinor into the other, hence:
\begin{equation}
C=\|\utilde{z}\|^2-\|z\|^2=0.\label{armatch}
\end{equation}
This constraint imposes, that the length of the bone is the same from whatever side we look at it, we call it the \emph{length matching constraint}. We can now invert equation \eref{spinrel} thus providing a parametrisation of the holonomy in terms of the spinors:
\begin{equation}
h=\frac{\E^{-\I\Phi}\ket{\utilde{z}}\bra{z}+\E^{\I\Phi}\roket{\utilde{z}}\robra{z}}{\|\utilde{z}\|\,\|z\|}\equiv\ou{h}{A}{B}=\frac{\E^{-\I\Phi}\utilde{z}^Az^\dagger_B+\E^{-\I\Phi}\utilde{z}^A_\dagger z_B}{\|z\|\,\|\utilde{z}\|}.\label{holpar}
\end{equation}
So far we have just described a way to parametrise both holonomy and flux by a pair of spinors, but the spinorial formalism extends further. It can also capture the Poisson algebra of $T^\ast SU(2)$. The symplectic structure for two pairs of harmonic oscillators 
\begin{equation}
\big\{z^\dagger_A,z^B\big\}=\frac{\I}{\hbar}\delta^B_A,\quad 
\big\{\utilde{z}^\dagger_A,\utilde{z}^B\big\}=-\frac{\I}{\hbar}\delta^B_A,\label{sympstruct}
\end{equation}
induce commutation relations for holonomies \eref{holpar} and fluxes \eref{fluxpar}:
\begin{subalign}
\big\{\ou{h}{A}{B},\ou{h}{C}{D}\big\}&=+\frac{2}{\hbar\ellp}\|z\|^{-4}\|\utilde{z}\|^{-2}C\epsilon^{AC}\ell_{BD}-
\frac{2}{\hbar\ellp}\|\utilde{z}\|^{-4}\|z\|^{-2}C\epsilon_{BD}\utilde{\ell}^{AC},\\
\{\ell_i,\ou{h}{A}{B}\}&=+\frac{\ellp}{\hbar}\ou{h}{A}{C}\ou{\tau}{C}{Bi},\\
\{\ell_i,\ell_j\}&=+\frac{\ellp}{\hbar}\uo{\epsilon}{ij}{m}\ell_m.
\end{subalign}
On the constraint hypersurface $C=0$ of the length matching constraint \eref{armatch}, these commutation relations reduce to the symplectic structure of $T^\ast SU(2)$, as given in \eref{holfluxal}. The Hamiltonian vector field $\mathfrak{X}_C=\{C,\cdot\}$ generates a flow inside the constraint hypersurface that leaves both holonomy \eref{holdef} and flux \eref{fluxdef} unchanged:
\begin{equation}
\exp(\varphi\hbar\mathfrak{X}_C)z^A=\E^{-\I\varphi}z^A,\quad\exp(\varphi\hbar\mathfrak{X}_C)\utilde{z}^A=\E^{-\I\varphi}\utilde{z}^A.
\end{equation}
Performing a symplectic quotient, thus projecting the orbits generated by $C$ into a point, we arrive at almost all of the original phase space, exempt only of the submanifold $T_o:=\{(h,X)\in SU(2)\times\mathfrak{su}(2)\simeq T^\ast SU(2)|X=0\}$ of vanishing flux, where we reach a coordinate singularity. In other words $(\C^2\times \C^2)/\!/_C=T^\ast SU(2)-T_o$.
\section{A spinorial action for discretised gravity}\label{newactn}
\subsection{Discretisation and partial continuum limit}\noindent
The last section studied the kinematical structure of three-dimensional Euclidean gravity on a simplicial lattice. Now, we introduce the dynamics of the theory as derived from an action variation. This action is the key novelty of the paper, and is based on what has been developed for the 3+1-dimensional case in \cite{hamspinfoam}. The derivation starts from a simplicial discretisation of the action \eref{actn}, but eventually yields again a continuum theory. This is possible through a partial continuum limit. The resulting action is a one-dimensional integral over the edges of the discretisation. The three-dimensional action integral thus turns into a sum over one-dimensional line integrals.

We start with the discretisation of the action \eref{actn} over the simplicial complex. This can be done with remarkable ease \cite{thiemann} and yields a sum over \emph{wedges}:
\begin{align}
S_M[e,A]&=-\frac{\hbar}{\ellp}\int_M e_i\wedge F^i\approx
-\frac{\hbar}{\ellp}\sum_{w:\text{wedges}}\int_{b_w}e_i\,\int_wF^i.\label{step1}
\end{align}
Here, we have split every spinfoam face $f$ into a sum over wedges, $f=\bigcup_{i=1}^Nw_i$. Figure \ref{tetra} gives an illustration of the geometry: A wedge $w$ \cite{Reisenberger:1996ib} is a triangular surface lying inside a spinfoam face $f$, two of its corners rest on an edge, the third belongs to the bone $b_w$ dual to $w$.  Both $b_w$ and $w$ carry an orientation that agrees with the orientation of $M$: If the pair of tangent vectors $(X,Y)$ is positively oriented in $w$, and $Z$ is positively oriented in $b_w$, the triple $(X,Y,Z)$ is positively oriented in $M$. The sum goes over only one of the two possible orientations of $b_w$. 

The main idea  of this section is to study a limiting process where the number of wedges goes to infinity. The result will turn the sum into an integral and give us a continuous action on each spinfoam face.
For the moment let us only study one particular wedge $w_o$ appearing in this sum. We then take the $SU(2)$ holonomy-flux variables and use them to parametrise the discretised action. For the flux the situation is immediate, looking back\footnote{Equation \eref{fluxdef} contains additional holonomies, here we have dropped them to keep our formulae simple, adding them, would not affect our final result.} at \eref{fluxdef} we trivially have:
\begin{equation}
\int_{b_{w_o}}e=\ell[b_{w_o}].\label{step2}
\end{equation}

For the second piece, the curvature term $\int_{w_o}F$ in the action, we use the holonomy as an approximation. Consider first the differential of the holonomy under variations of the underlying path. Let  $\gamma_\varepsilon:[0,1]\rightarrow M,s\mapsto\gamma_\varepsilon(s)$ be an  $\varepsilon$-parameter family of paths. Taking derivatives with respect to $s$ and $\varepsilon$ we obtain the tangent vectors $\delta\gamma_\varepsilon(s)=\frac{\di}{\di\varepsilon}\gamma_\varepsilon(s)\in T_{\gamma_\varepsilon(s)}M$ and  $\gamma^\prime_\varepsilon(s)=\frac{\di}{\di s}\gamma_\varepsilon(s)\in T_{\gamma_\varepsilon}(s)M$.  Simplifying our notation we write $\delta\gamma\equiv\delta\gamma_{\varepsilon=0}$ and equally for all other quantities at $\varepsilon=0$. We can then find the variation of the holonomy directly from its defining differential equation:
\begin{equation}
\frac{\di}{\di s}h_{\gamma_\varepsilon(s)}=-A_{\gamma_\varepsilon(s)}(\gamma_\varepsilon^\prime)\,h_{\gamma_\varepsilon(s)}.\label{defeq}
\end{equation}
This works as follows: We just take the differential of \eref{defeq} with respect to $\varepsilon$, multiply everything by $h_{\gamma_\varepsilon(s)}^{-1}$ and integrate the resulting equation from $s=0$ to $s=1$. A partial integration eventually yields the desired variation of the holonomy:
\begin{equation}
\frac{\di}{\di \varepsilon}\Big|_{\varepsilon=0}h_{\gamma_\varepsilon(1)}=-A_{\gamma(1)}(\delta\gamma)h_{\gamma(1)}+h_{\gamma(1)}A_{\gamma(0)}(\delta\gamma)+
\int_0^1\di sh_{\gamma(1)}h_{\gamma(s)}^{-1}F_{\gamma(s)}(\gamma^\prime,\delta\gamma)h_{\gamma(s)}.\label{step3}
\end{equation}

\begin{figure}
     \centering
     \includegraphics[width= 0.4\textwidth]{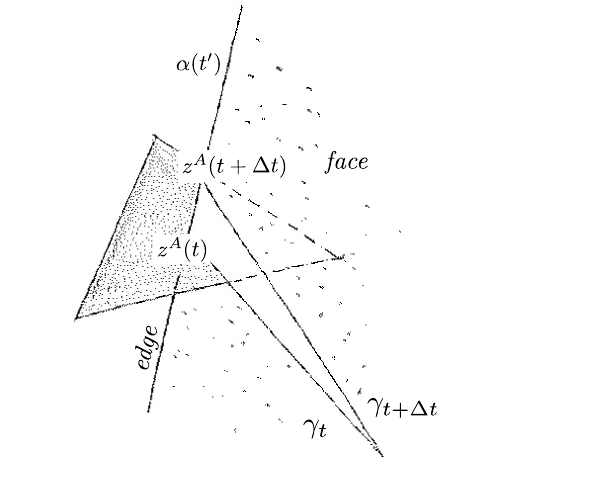}
     \caption{Going from $t$ to $t+\Delta t$ we can probe an infinitesimal wedge, the boundary of which has two parts. The first part belongs to the edge and has a tangent vector $\dot{\alpha}$. The second part (the triangular line in the picture) is a link inside the  face, and splits into two halves. Its \qq{upper} part we call $\gamma_{t+\Delta t}$, while the lower half is $\gamma_{t}$, putting them together determines $z^A(t+\Delta t)$: The spinor $z^A(t+\Delta t)$ is the parallel transport of $z^A(t)$ along the connecting link $\gamma_{t+\Delta t}^{-1}\circ\gamma_t$ modulo an overall phase $\Phi(t+\Delta t)$.}
     \label{inft}
\end{figure}

Consider now the boundary $\partial f$ of the underlying spinfoam face. This is a one-dimensional loop $\alpha:[0,1]\rightarrow M, t\mapsto \alpha(t)$, parametrised by some $t\in[0,1]$. The boundary of the wedge touches this loop in a small segment $\alpha(t_o,t_o+\Delta t)\subset\partial w_o$ corresponding to some interval $[t_o,t_o+\Delta t]$ in $t$. Two more sides bound the wedge $w_o$, these are the \emph{half links} $\gamma_{t_o}$ and $\gamma_{t_o+\Delta t}$: The path $\gamma_t\subset f$ connects the point $\alpha(t)$ on the boundary of $f$ with the bone dual to the face: $\gamma_t(0)=\alpha(t)$ and $\gamma_t(1)=b_{w_o}\cap f$. Figure \ref{tetra} and \ref{inft} should further clarify the situation. 

Next, we take the holonomy $h_{\gamma_t}=\mathrm{Pexp}(-\int_{\gamma_t}A)$ along the connecting link and study its velocity as we move forward in $t$. This gives the infinitesimal change of $h_{\gamma_t}$ under a variation $\gamma_t\rightarrow\gamma_t+\varepsilon\delta\gamma_t$ of the underlying path---a derivative just as in \eref{step3}. The variation of the path vanishes at $t=1$, because all paths meet at the center of the spinfoam face: $\forall t,t^\prime\in[0,1]:$ $\gamma_t(1)=\gamma_{t^\prime}(1)$, thus: $\frac{\di}{\di t}\gamma_t(1)=0$, hence:
\begin{equation}
\frac{\di}{\di t}h_{\gamma_t(1)}-h_{\gamma_t(1)}A_{\gamma_t(0)}\Big(\frac{\di}{\di t}\gamma_t\Big)=\int_0^1\di s\,h_{\gamma_t(1)}h^{-1}_{\gamma_t(s)}
F_{\gamma_t(s)}\Big(\frac{\di}{\di s}\gamma_t(s),\frac{\di}{\di t}\gamma_t(s)\Big)h_{\gamma_t(s)}.
\end{equation}
We can now use this equation to write the smeared curvature tensor as the covariant time derivative of the link holonomy: 
\begin{equation}
h_{\gamma_t(1)}^{-1}\frac{D}{\di t}h_{\gamma_t(1)}\equiv
h_{\gamma_t(1)}^{-1}\Big(\frac{\di}{\di t}h_{\gamma_t(1)}-h_{\gamma_t(1)}A_{\alpha(t)}(\dot\alpha)\Big)=\int_0^1\di s\,h^{-1}_{\gamma_t(s)}
F_{\gamma_t(s)}\Big(\frac{\di}{\di s}\gamma_t(s),\frac{\di}{\di t}\gamma_t(s)\Big)h_{\gamma_t(s)}.\label{holvar}
\end{equation}

Let us now isolate the contribution $S_{w_o}$ to the discretised action \eref{step1} coming from the wedge $w_o$. Inserting our curvature formula \eref{holvar} into the discretised action \eref{step1} we find that each wedge adds the term
\begin{equation}
S_{w_o}\approx-\frac{2\hbar\Delta t}{\ellp}\,\ell_{AB}[b_{w_o}]\Big(h_{\gamma_{t_o}(1)}^{-1}\frac{D}{\di t}h_{\gamma_{t_o}(1)}\Big)^{AB}\label{step4}
\end{equation} to the total action \eref{step1}.
This approximation improves as the wedge shrinks to a line, where it becomes exact.
For the flux $\ell_{AB}[b_{w_o}]$, we can now find a diagonalising spinor $z^A$ just as in equation \eref{fluxpar} above. Since this spinor belongs to the frame at $t=t_o$, its better to write $z^A=z^A(t_o)$, and we get:
\begin{equation}
\ell_{AB}[b_{w_o}]=\frac{\ellp}{2\I}z_{(A}(t_o)z^\dagger_{B)}(t_o).\label{step5}
\end{equation}
We can repeat this construction for all other values of $t$, thus obtaining a map $z^A:[0,1]\rightarrow \C^2, t\mapsto z^A(t)$. For each value of $t$, the spinor is unique up to an overall phase. We can always choose this phase such that the spinor $z^A(t)$ is continuous in $t$. It should also respect the periodicity of the underlying loop $\alpha$, hence $z^A(0)\stackrel{!}{=}z^A(1)$. 

We now turn to the link holonomy $h_{\gamma_t}$ connecting $\alpha(t)$ with the center of the spinfoam face $f$. We introduce an additional spinorial field $w^A:[0,1]\rightarrow\C^2,t\mapsto w^A(t)$ in the frame at the center $b_{w_o}\cap f$ of the face and use the pair $(z^A,w^A)$ of spinors to parametrise the connecting holonomy. Going back to \eref{holpar} we get the precise relation:
\begin{equation}
\big[h_{\gamma_t(1)}\big]^A_{\phantom{A}B}=\frac{w^A(t)z^\dagger_B(t)+w_\dagger^A(t)z_B(t)}{\|w(t)\|\,\|z(t)\|}.\label{step6}
\end{equation}
Just as $z^A(t)$ also $w^A(t)$ shall be both continuous and periodic in $t$: $w^A(0)=w^A(1)$. Compared to \eref{holpar} we have ignored the possibility of a relative phase $\Phi$ between $z^A$ and $w^A$. Setting $\Phi=0$ does not affect our final result. The spinors $z^A$ and $w^A$ are not independent, once again we must respect the length matching constraint \eref{armatch}:
\begin{equation}
C=\|w\|^2-\|z\|^2\stackrel{!}{=}0.\label{lengthmatch2}
\end{equation}

There is a subtlety with the covariant time derivative of these spinors: $z^A(t)$ belongs to the frame at $\alpha(t)$, while $w^A(t)$ is a spinor living at the center of the spinfoam face. The tangent vector $\frac{\di}{\di t}\gamma_t(s)$ vanishes at $s=1$ because $\gamma_t(s=1)$ is the same for all values of $t$---this is just a point in the center of the spinfoam face. On the other hand $\gamma_t(0)=\alpha(t)$, hence:
\begin{equation}
\frac{D}{\di t}z^A(t)=\dot{z}^A(t)+A^i_{\mu}\big(\alpha(t)\big)\dot{\alpha}^\mu(t)\ou{\tau}{A}{Bi}z^B(t),\quad\text{but}\quad
\frac{D}{\di t}w^A(t)=\dot{w}^A(t).\label{tderiv}
\end{equation}
Once again $\alpha(t)$ denotes the loop bounding the spinfoam face $f$, $\dot{\alpha}^\mu(t)$ is its tangent vector, while $\ou{A}{i}{\mu}$ are the $SU(2)$ connection components with respect to the canonical generators $\{\tau_i\}_{i=1,2,3}$ of $\mathfrak{su}(2)$ ($2\I\tau_i$ are the usual Pauli matrices).

Inserting the velocities \eref{tderiv} together with equations \eref{step6} and \eref{step5} into our expression for the wedge action \eref{step4} we eventually get:
\begin{equation}
S_{w_o}=\frac{\I\hbar\Delta t}{2}\left(
\frac{\|z\|^2}{\|w\|^2}w^\dagger_A\dot{w}^A-\frac{\|z\|^2}{\|w\|^2}w_A\dot{w}^A_\dagger
+z_A\frac{D}{\di t}z_\dagger^A-z^\dagger_A\frac{D}{\di t}z^A
\right)\Big|_{t=t_o}.
\end{equation}
Let us now repeat this construction for all wedges  $w_i$ appearing in the decomposition of the spinfoam face $f=\bigcup_{i=1}^Nw_i$. The discretisation should be uniform in $t$: We can always choose the $t$-coordinate such that the  $i$-th wedge $w_i$ intersects the boundary $\partial f$ in the $t$-interval $[\frac{i-1}{N},\frac{i}{N}]$. The difference $\Delta t$ thus represents the fraction $N^{-1}$. Sending $N\rightarrow \infty$ leads us to an integral over the entire spinfoam face:
\begin{equation}
S_f=\frac{\I\hbar}{2}\int_0^1\di t\left(
\frac{\|z\|^2}{\|w\|^2}w^\dagger_A\dot{w}^A-\frac{\|z\|^2}{\|w\|^2}w_A\dot{w}^A_\dagger
+z_A\frac{D}{\di t}z_\dagger^A-z^\dagger_A\frac{D}{\di t}z^A
\right).\label{step7}
\end{equation}
Functional variations of the spinors must respect the length matching constraint \eref{lengthmatch2}. We can account for this $C=0$ constraint by introducing a Lagrange multiplier $\varphi$, and adding the term $\varphi\,C$ to the action. Notice now, that on the constraint hypersurface $C=0$, the variation of the fraction $\|z\|^2/\|w\|^2$ turns into the variation of the constraint itself:
\begin{equation}
\delta\left(\frac{\|z\|^2}{\|w\|^2}\right)\Big|_{C=0}=-\frac{\delta C}{\|z\|^2}.
\end{equation}
Therefore, variations of $\|z\|^2/\|w\|^2$ just shift the value of the Lagrange multiplier $\varphi$. In other words, we can ignore these two fractions and work  with a simplified action instead:
\begin{align}
\nonumber S_f[z,w,\varphi,A]&=\frac{\I\hbar}{2}\int_0^1\di t\Big(
w^\dagger_A\dot{w}^A-w_A\dot{w}^A_\dagger
+z_A\frac{D}{\di t}z_\dagger^A-z^\dagger_A\frac{D}{\di t}z^A
+2\I\varphi\big(\|z\|^2-\|w\|^2\big)
\Big)=\\
&=-\I\hbar\int_0^1\di t\Big(z^\dagger_A\frac{D}{\di t}z^A-w^\dagger_A\dot{w}^A-\I\varphi\big(\|z\|^2-\|w\|^2\big)\Big).\label{step8}
\end{align}
The last step involved a partial integration, which, thanks to the periodicity of the spinors, does not yield any additional boundary terms.

Each spinfoam face contributes through equation \eref{step8} to the total action \eref{step1}. Equation \eref{step8} is a functional that depends on three elements: the spinors $z^A$ and $w^A$, the gauge connection $A$, and a $U(1)$ angle $\varphi$. Let us now repeat the construction for the entire simplicial decomposition. We thus have spinor fields $z^A_f:\partial f\rightarrow \C^2$, $w^A_f:\partial f\rightarrow \C^2$, and a $U(1)$ angle $\varphi_f:\partial f\rightarrow\R$ attached to each face $f$. The $SU(2)$ connection $A_e(t)=A_\mu(e(t))\dot{e}^\mu(t)\in\mathfrak{su}(2)$ belongs to the edges $e$ of the discretisation, where $\dot{e}(t)\in T_{e(t)}M$ denotes the corresponding tangent vector. The boundary conditions are such, that all spinors are continuous once we go around the spinfoam face (the angle $\varphi_f$ must only be periodic modulo $2\pi$). We thus get the following action for the discretised manifold $M$: 
\begin{align}\nonumber
S_M[z_{f_1},z_{f_2},&\dots;w_{f_1},w_{f_2},\dots;\varphi_{f_1},\varphi_{f_2},\dots;A_{e_1},A_{e_2},\dots]=\\
\nonumber&\equiv S_M[\underline{z},\underline{w},\underline{\varphi},\underline{A}]=-\I\hbar\sum_f\oint_{\partial f}\Big(z^\dagger_{fA} D\,z^A_f-w^\dagger_{fA}\di w^A_f-\di t\,\I\varphi_f\big(\|z_f\|^2-\|w_f\|^2\big)\Big)=\\
&=-\I\hbar\sum_f\oint_{\partial f}\Big(\langle z_f| D|z_f\rangle-\langle w_f|\di|w_f\rangle-\di t\,\I\varphi_f\big(\|z_f\|^2-\|w_f\|^2\big)\Big).\label{step9}
\end{align}
The action does not care about the value of the connection in the \qq{bulk}, it only probes the connection along the edges through the covariant $t$-derivative:
\begin{equation}
\frac{D}{\di t}z^A(t)=\dot{z}^A(t)+A^i_e(t)\ou{\tau}{A}{Bi}z^B(t),\quad\text{with}:\quad A^i_e(t)=\ou{A}{i}{\mu}\big(e(t)\big)\dot{e}^\mu(t),\label{edgeconn}
\end{equation}
where $t$ parametrises the edge $e$, and $\dot{e}^\mu(t)$ denotes its tangent vector.

\subsection{Equations of motion and gauge symmetries}
\noindent Now that we have a continuous action \eref{step9} for the discretised manifold we have to find its extremum, identify the equations of motion and compare their solutions with those \eref{conteom} of the continuum theory. All fields in the action \eref{step9}---the spinors, the gauge connection $A^i$, and the Lagrange multipliers $\varphi$---have support only on the one-dimensional edges of the discretisation. The resulting equations of motion are therefore all local in $t$. This is a huge simplification 
compared to other discretisation schemes, where one has to deal with difference equations instead (see for instance: \cite{Dittrich:2013jaa, Hoehn:2014fka, Hoehn:2014wwa}). Here, all fields are continuous along the edges of the discretisation.

Our analysis of the action variation splits into four steps: First of all, we give the evolution equations along the edges of the discretisation, we then study the constraint equations, and eventually speak about the canonical formalism and the gauge symmetries of the theory.

\emph{(i) Evolution equations.} We start with the evolution equations for the spinors. The spinor fields $z^A_f$ and $w^A_f$ only appear in the corresponding face action $S_f[z_f,w_f,\varphi_f,A]$. Its action variation yields the evolution equations:
\begin{equation}
\frac{D}{\di t}z^A=\I\varphi z^A,\quad\text{and}\quad \frac{\di}{\di t}w^A=\I\varphi w^A,\label{evolveq}
\end{equation}
where we have dropped the face label $z^A_f\equiv z^A$ for simplicity. We can immediately integrate these equations. The holonomy parallel transports the $z$-spinors up to an overall phase, and this phase also turns the $w$-spinors around:   
\begin{equation}
z^A(t)=\E^{\I\Phi(t)}\ou{U}{A}{B}(t)z^B(0),\quad w^A(t)=\E^{\I\Phi(t)}w^A(0).\label{evolvsol}
\end{equation} 
We have introduced some new elements here: $U(t)$ is the $SU(2)$ holonomy around the boundary of the spinfoam face, from $t_0=0$ to $t_1=t$, while the integral over the Lagrange multiplier $\varphi$ gives the overall angle $\Phi(t)$:
\begin{equation}
\ou{U}{A}{B}(t)=\mathrm{Pexp}\Big(-\int_0^t\di sA_\mu\big(\alpha(s)\big)\dot{\alpha}^\mu(s)\Big)^A_{\phantom{A}B},\quad\text{and}\quad
\Phi(t)=\int_0^t\di s\varphi(s).\label{spinsol}
\end{equation}
Furthermore, $\alpha:[0,1]\rightarrow M$ bounds the spinfoam face $f$, and the orientation of $\alpha$ agrees with the induced orientation of $\partial f$. 

Let us now see what happens once we go around the spinfoam face and close 
$\alpha(t)$, hence forming a loop. The $w^A$-spinors are periodic in $t$, looking back at (\ref{evolvsol}, \ref{spinsol}), we see this immediately implies that:
\begin{equation}
\forall f:\exists\, n_f\in\Z:\int_{\partial f}\di t\varphi_f=2\pi n_f.\label{windnumb}
\end{equation}
This, together with the periodic boundary conditions for the $z^A$-spinors gives us an eigenvalue equation for the $SU(2)$-holonomy around the bounding loop:
\begin{equation}
z^A(0)=\ou{U}{A}{B}(1)z^B(0).
\end{equation}
Having one twice degenerate eigenvalue, this $SU(2)$ element can only be the identity: $U(1)=\mathds{1}$. This must be true for all spinfoam faces appearing in the simplicial complex, hence:
\begin{equation}
\forall f: \ou{h}{A}{B}[\partial f]:=\mathrm{Pexp}\Big(-\oint_{\partial f}A\Big)^A_{\phantom{A}B}=\delta^A_B.\label{flatconst}
\end{equation}
This is the discrete analogue of the flatness condition $\ou{F}{i}{\mu\nu}=0$, i.e. equation \eref{flatcons}, because the holonomy well approximates the curvature in the spinfoam face $f$: $h^{(AB)}[\partial f]\approx \int_fF^{AB}$. We have thus recovered already one half of the equations of motion \eref{conteom} of the continuum theory. What about the other half, that is the vanishing of torsion $D_{[\mu}\ou{e}{i}{\nu]}=0$ as implied by equation \eref{torscons}? In the continuum, the vanishing of torsion follows from the connection variation. The same happens in the discrete: The variation of the spinor action \eref{step9} with respect to the $SU(2)$ connection $A_e(t)$ on the edges $e$ will give us the discrete version of the torsionless condition. This is the Gauß law, which brings us to our analysis' next step: 
 
\emph{(ii) Constraint equations.} We now study the constraint equations of the theory, and we start with the Gauß law. We obtain it from the variation of the action \eref{step9} with respect to the gauge potential $A^i_e(t)$ on the edges (as defined in \eref{edgeconn} as the $SU(2)$ connection contracted with the tangent vector $\dot{e}^\mu$). This gauge potential only appears in the covariant derivative $D$ of the $z$-spinors into the direction of the edge. Every edge bounds three faces $f$, each of which carries its own $z_f$-spinor. There are thus three such differentials $D/\di t\,z^A_f$ for each value of $t$. Since we also have the $w_f$-spinors, there are altogether six spinors per edge. To keep our notation simple, let us only study one edge $e$ in the triangulation, and call the corresponding spinors $(z^A_f,w^A_f)_{f=1,2,3}$, where $f=1,2,3$ label the three adjacent faces. 
The edge $e$ thus contributes to the full action \eref{step9} for the discretised manifold $M$ through the expression:
\begin{align}\nonumber
S_e[\uline{z},\uline{w},{\uline{\varphi}},A_e]&=-\I\hbar\sum_{f=1}^3\int_{t_o}^{t_1}\di t\Big(z^\dagger_{fA}\frac{\di}{\di t}z^A_f+z^\dagger_{fA}\ou{\tau}{A}{Bi}\,
z^B_f\,A^i_e(t)-w^\dagger_{fA}\frac{\di}{\di t}w^A_f-\I\varphi_f\big(\|z_f\|-\|w_f\|^2\big)\Big)=\\
&\equiv-\I\hbar\sum_{f=1}^3\int_{t_o}^{t_1}\di t\Big(\big\langle z_f\big|\frac{\di}{\di t}\big|z_f\big\rangle+\langle z_f|\tau_i|z_f\rangle A^i_e(t)-\big\langle w_f\big|\frac{\di}{\di t}\big|w_f\big\rangle-\I\varphi_f\big(\|z_f\|-\|w_f\|^2\big)\Big).
\label{eactn}
\end{align}
which we shall call the \emph{edge action}.
Note, that an implicit assumption is hiding here: There are three spinfoam faces meeting at the edge $e$ (again we refer to figure \ref{tetra} for an illustration), and each of them carries an orientation. These orientations may not match the orientation of the edge $e$, while in equation \eref{eactn} we have implicitly assumed so. Did the orientations not match, relative sign factors would be necessary. We could then, however, always absorb those factors of $\pm 1$ into a redefinition of the spinors: The replacement $z^A\rightarrow z^A_{\dagger}$ would bring us back to \eref{eactn}, modified only by a boundary term that is irrelevant for the following argument.

Variation of \eref{eactn} with respect to the connection $A^i_e$ gives us a constraint: The three internal vectors $\ell_i[b]$, defined as in \eref{step5}, 
must close to form a triangle:
\begin{equation}
G_i:=\frac{\hbar}{\ellp}\sum_{f=1}^3\ell_i[b_f]_t=\I\hbar\sum_{f=1}^3\ou{\tau}{AB}{i} z^\dagger_{fA}(t)z_{fB}(t)=0.\label{gausslaw}
\end{equation}
The vanishing of $G_i$ has a clean geometric interpretation. It gives us a discretisation of the torsionless equation $D_{[\mu}\ou{e}{i}{\nu]}=0$, i.e. equation \eref{conteom}, smeared over the triangle dual to the edge. Indeed, there is the non-Abelian version of Stoke's theorem, and it tells us that for any triangle $\tau$ the fluxes through its bounding sides sum up to zero: $\int_\tau T^i=\int_\tau De^i=\int_{\partial \tau}e^i=\sum_{b\subset \partial T}\int_be^i=\sum_{b\subset \partial T}\ell^i[b]$. This series of equations is true only in a small neighbourhood, where we can map the internal $\mathfrak{su}(2)$ index $i$ into a common frame, reached by a family of holonomies just as in equation \eref{fluxdef}. The geometric interpretation of \eref{gausslaw} is immediate; for, the vector ${\ell}^i[b]_t\in\R^3$ represents the bone $b$ in the internal frame at the point $e(t)$ of the edge. 

The three bounding sides close to form a triangle, but this is not any triangle, it is the triangle dual to the edge mapped into the local frame of reference. 
As we go along the edge, and move forward in $t$ this triangle preserves its shape, the evolution equations \eref{eactn} for the spinors $z^A_f(t)$ just rigidly turn it around.

There is one more constraint to be studied. The variation of the Lagrange multiplier $\varphi$ yields the length matching condition \eref{lengthmatch2}. Once again its geometrical meaning is immediate. The constraint $C=0$ tells us that the $w$-spinors and $z$-spinors describe the very same geometrical object, the triad smeared over the bounding bones, evaluated just in two different frames, one at the center of the spinfoam face, the other attached to its boundary. There is a unique $SU(2)$ element \eref{step6} that maps one of these spinors into the other, and it gives us the parallel transport from the boundary of the spinfoam face towards its center. 

\emph{(iii) Canonical formalism.} Looking at the edge action, we can immediately read off the symplectic structure. The elementary Poisson brackets are:
\begin{equation}
\big\{z^\dagger_{fA},z^B_{f^\prime}\big\}=+\frac{\I}{\hbar}\delta_{ff^\prime}\delta^B_A,\quad
\big\{w^\dagger_{fA},w^B_{f^\prime}\big\}=-\frac{\I}{\hbar}\delta_{ff^\prime}\delta^B_A,\label{symplstruct}
\end{equation}
while all mutual Poisson brackets between the $w$- and $z$-spinors vanish. Notice, that this agrees with our conventions from our introductory section \ref{spinrep}. Equation \eref{sympstruct} introduced the spinorial Poisson brackets essentially by hand, here they naturally fall out of the formalism. 

The evolution equations \eref{evolveq} are generated by a Hamiltonian. This $t$-dependent Hamiltonian is a sum over both Gauß's law and the triple of length matching constraints:
\begin{equation}
H_t=-\I\hbar\sum_{f=1}^3\Big[A_e(t)^i\ou{\tau}{AB}{i}\,z^\dagger_{fA}z_{fB}+\I\varphi_f\big(\|z_f\|^2-\|w_f\|^2\big)\Big].\label{edgeham}
\end{equation}
That the Hamiltonian is a sum over constraints, and hence vanishes, should not surprise us. Indeed, the action is a prototypical example of a timeless systems \cite{rovelli}, invariant under reparametrisations in $t$. We are thus dealing with a general covariant system, systems for which the Hamiltonian always turns into a sum over constraints. 

Although the Hamiltonian vanishes, this does not mean that the evolution equations are totally trivial. If $F:(\C^2\times\C^2)^3\rightarrow \C, (z^A_f,w^A_f)_{f=1,2,3}\mapsto F[(z^A_f,w^A_f)_{f=1,2,3}]$ is a function on the phase space of an edge, the Hamilton equations imply, in fact:
\begin{equation}
\frac{\di}{\di t}F_t=\big\{H,F\big\}_t=\mathfrak{X}_H[F]_t\stackrel{\text{in general}}{\neq} 0.\label{hameq}
\end{equation}

The action \eref{eactn} describes twelve harmonic oscillators coupled by six first-class constraints. The first three of them \eref{gausslaw} impose the vanishing of the total \qq{angular momentum} of the system, the other three---these are the length matching conditions \eref{lengthmatch2} on the faces---require that the spinors have equal \qq{energy}: $C_f=\|z_f\|^2-\|w_f\|^2=0$. This \qq{energy condition} resonates with recent developments of Frodden, Gosh and Perez \cite{FGPfirstlaw} and Bianchi \cite{Bianchientropy}, who argued that in four dimensions the horizon area measures the local energy of a stationary observer at short distance from the horizon. In three dimensions area becomes length, and indeed the length $\boldsymbol{L}[b]=\ellp\|w\|^2$ of the bones $b$ linearly appear in our edge Hamiltonian \eref{edgeham}. At the moment, this analogy is very vague, and deserves a more profound investigation. 

\emph{(iv) Gauge symmetries.} What are the gauge symmetries of the system? First of all, there is the one-dimensional diffeomorphism invariance of the action. Replacing the $t$-coordinate by $\tilde{t}(t)$, leaves the action invariant, provided we also change the Lagrange multipliers appropriately:
\begin{equation}
\tilde{A}_e^i(\tilde{t})=\frac{\di t}{\di \tilde{t}}A^i_e(t),\quad\text{and}\quad \tilde{\varphi}(\tilde{t})=\frac{\di t}{\di \tilde{t}}\varphi(t).
\end{equation}
This gives us the first gauge symmetry. Then, there are those symmetries that are generated by the Hamiltonian vector field of the constraints of the system: In fact, the length matching constraint generates $U(1)$ gauge transformations:
\begin{subequations}
\begin{align}
\tilde{z}^A(t)=\E^{-\I\lambda(t)}z^A(t)&=\exp\big(\lambda(t)\hbar\mathfrak{X}_C\big)z^A\big|_t,\\
\tilde{w}^A(t)=\E^{-\I\lambda(t)}w^A(t)&=\exp\big(\lambda(t)\hbar\mathfrak{X}_C\big)w^A\big|_t,
\end{align}\label{u1trafos}
\end{subequations}
where $\mathfrak{X}_C=\{C,\cdot\}$ is the Hamiltonian vector field of the constraint. These generators transform each $(z,w)$-pair of spinors independently, so we rather have an $U(1)^3$ symmetry per edge. The equations \eref{u1trafos} alone would not preserve the Lagrangian \eref{eactn}. The $U(1)$ gauge symmetry also shifts the Lagrange multipliers, which transform as $U(1)$ gauge potentials:
\begin{equation}
\tilde{\varphi}_f(t)={\varphi}_f(t)+\dot{\lambda}_f(t).\label{lambdatrafo}
\end{equation}

The internal $SU(2)$ invariance gives us another local gauge symmetry. The Gauß constraint $G_i$ generates, in fact, local $SU(2)$ gauge transformations $g(t)\in SU(2)$, and rigidly moves the spinors around:
\begin{equation}
\tilde{z}^A_f(t)=\exp\big(-\Lambda^i(t)\mathfrak{X}_{G_i}\big)\,z^A_f\big|_t=\ou{g^{-1}(t)}{A}{B}\,z^B_f(t),\quad\text{with}\quad g(t)=\exp\big(\Lambda^i\tau_i\big).\label{gaugetrafo1}
\end{equation}
Just as for the $U(1)$ symmetry, we also have to change the gauge potential to keep the action invariant. The gauge potential $A^i_e(t)$ defines a $SU(2)$ connection on a line, hence transforms inhomgenously under $SU(2)$:
\begin{equation}
\tilde{A}_e(t)=\big(\rho_\Lambda{A}\big)_e(t):=g^{-1}(t)\frac{\di}{\di t}g(t)+g^{-1}(t)A_e(t)g(t).\label{gaugetrafo2}
\end{equation} 
The local $SU(2)$ transformations \eref{gaugetrafo1}, together with \eref{gaugetrafo2} clearly preserve the Lagrangian \eref{eactn}. In summary, the action \eref{eactn} has three local gauge symmetries: First of all, there is the reparametrisation invariance in $t$, next there are $U(1)$ phase transformations for each individual pair of $(z_f,w_f)$-spinors, and then there are also $SU(2)$ transformations for the triple of spinors $(z_f)_{f=1,2,3}$ on an edge. These $SU(2)$ rotations move the dual  triangle in internal space, but preserve its overall shape. 
\section{Path-integral quantisation}\label{pathintsec}
\noindent We are now ready to study the resulting quantum theory, and define the vacuum to vacuum amplitude $\langle \Omega|\Omega\rangle=Z_M$ for the discretised manifold $M$ as the path integral over the exponential of the spinorial action \eref{step9}, and hence study the following expression:
\begin{equation}
Z_M=\smashoperator{\int_{\substack{\text{all spinors be}\\ \text{periodic in $\partial f$}}}}\qquad\prod_{f:\text{faces}}\mathcal{D}[z_f]\mathcal{D}[w_f]\mathcal{D}[\varphi_f]\Delta_{\mathrm{FP}}^\psi[\varphi_f]\delta\big(\psi[\varphi_f]\big)
\prod_{e:\text{edges}}\mathcal{D}[A_e]\Delta_{\mathrm{FP}}^\Psi[A_e]\delta\big(\Psi[A_e]\big)\E^{\frac{\I}{\hbar}S_M[\underline{z},\underline{w},\underline{\varphi},\underline{A}]}\label{pathint}.
\end{equation}
The underlying manifold $M$ shall be closed, and the amplitude $Z_M$ is therefore a pure number, not depending on any boundary data. Insertions of gauge invariant observables give the $n$-point functions of the theory.
All fields are supported only on the one-dimensional edges of the discretisation, and fulfil periodic boundary conditions once we go around a spinfoam face. Furthermore, $\prod_f\mathcal{D}[z_f]$ with $\mathcal{D}[z]=\prod_t\frac{d^4z(t)}{\pi^2}$ denotes the flat integration measure in the infinite dimensional space of spinor-valued functions over the edges of the discretisation.\footnote{Equally for $\mathcal{D}[A_e]$, and $\mathcal{D}[\varphi_f]$: They are formal Lebesgue measures in the space of $\mathfrak{su}(2)$-valued functions $A_e:e\rightarrow\mathfrak{su}(2)$ and real-valued functions $\varphi_f:\partial f\rightarrow\R$ respectively.} The spinorial action $S_M$ \eref{step9} has local $U(1)$ and $SU(2)$ gauge symmetries (\ref{u1trafos}--\ref{gaugetrafo2}) on the edges. This necessitates a gauge fixing, for we cannot integrate over the 
gauge orbits, because this generically yields an infinity. We thus take the integral only over a gauge fixing surface, which intersect every gauge orbit exactly once. The gauge fixing functions $\Psi[A]$ and $\psi[\varphi]$ define such a gauge section, while the corresponding Faddeev--Popov determinants, $\Delta^\Psi_{\mathrm{FP}}[A]$ and $\Delta^\psi_{\mathrm{FP}}[\varphi]$ are needed to end up with a gauge invariant integration measure. This also guarantees the invariance of the resulting amplitude under small deformations of the gauge fixing surface.

\emph{Step 0: Bargmann's quantisation of the harmonic oscillator.} Before we go on and actually calculate this integral, let us first study the kinematical structure of the resulting quantum theory, its Hilbert space and operators \cite{Livinerep, Dupuis:2012vp}. We start from the space of analytic functions $\mathcal{H}\in f:\C^2\rightarrow \C, z\mapsto f(z)$, which carry a natural representation of the classical commutation relations $\{z^\dagger_A,z^B\}=\frac{\I}{\hbar}\delta^B_A$. Following Bargmann's analytic quantisation of the harmonic oscillator, the $z^A$-spinor acts by multiplication, while ${z}^\dagger_A$ turns into a derivative:
\begin{equation}
\big(\hat{z}^Af\big)(z)=z^Af(z),\quad \big(\hat{z}^\dagger_Af\big)(z)=\frac{\partial}{\partial z^A}f(z).
\end{equation}  
 The reality conditions ${\bar{z}}^{\bar A}=\delta^{A\bar A}{z}^\dagger_A$ uniquely determine the inner product as the Gaußian integral:
\begin{equation}
\big\langle f, f^\prime\big\rangle=\frac{1}{\pi^2}\int_{\C^2}d^4z\E^{-\delta_{A\bar A}z^A\bar{z}^{\bar A}}\overline{f(z)}f^\prime(z),\label{innprod}
\end{equation}
where $d^4z=-\frac{1}{4}\di z^0\di\bar{z}^{\bar 0}\di z^1\di\bar{z}^{\bar 1}$ is the flat integration measure, and both $f$ and $f^\prime$ are analytic in $\C^2$. A short moment of reflection reveals the Gaußian measure $\propto d^4z\exp(-\|z\|^2)$ truly respects the reality conditions: $\forall f,f^\prime\in\mathcal{H}:$ $\big\langle f, \hat{z}^Af^\prime\big\rangle=\big\langle \hat{\bar{z}}^{\bar A}f, f^\prime\big\rangle$. Next, we need a complete orthonormal basis. A convenient choice is given by the following family of polynomials:
\begin{equation}
\langle z|j,m\rangle=\frac{1}{\sqrt{(j-m)!(j+m)!}}\big(z^0\big)^{j-m}\big(z^1\big)^{j+m},\quad\text{and}\quad
\langle j,m|j^\prime,m^\prime\rangle=\delta_{jj^\prime}\delta_{mm^\prime}.\label{canbasis}
\end{equation} 

Then there are the operators. The quantisation of the fluxes \eref{fluxdef} yields the generators of angular momentum:
\begin{equation}
L_i=\I\tau^{AB}_{\phantom{AB}i}\hat{z}^{\phantom{\dagger}}_A\hat{z}^\dagger_B=-\ellp^{-1}\hat{\ell}_i,\label{su2gen}
\end{equation}
which satisfy the usual commutation relations $[L_i,L_j]=\I\uo{\epsilon}{ij}{m}L_m$ of $\mathfrak{su}(2)$. Another important operator is the spinor's norm. Any homogenous function diagonalises this operator, choosing a normal ordering we find, in fact:
\begin{equation}
{:}\|\hat{z}\|^2{:}=\frac{1}{2}\big(\hat{z}^A\hat{z}^\dagger_A+\hat{z}^\dagger_A\hat{z}^A\big)=z^A\frac{\partial}{\partial z^A}+1,\quad\text{thus}\quad
{:}\|\hat{z}\|^2{:}\big|j,m\big\rangle=(2j+1)\big|j,m\big\rangle.\label{normop}
\end{equation}
This gives us the spectrum of the length operator: Classically, each bone has a physical length given by $\boldsymbol{L}[b]=\sqrt{\ell_i[b]\ell^i[b]}$, but now the $z$-spinors parametrise the fluxes \eref{fluxpar}, and their squared norm measures the length of $b$. Choosing a normal ordering, and looking back at \eref{normop} we thus get the spectrum of the length operator:
\begin{equation}
\mathrm{spec}\big(\hat{\boldsymbol{L}}\big)=\big\{\ellp(j+\tfrac{1}{2})\big\}_{2j\in\N_0}.
\end{equation}

\emph{Step 1: Integration over the spinors.} The integral over the spinors factorises into a product over the individual spinfoam faces. We take the contribution from a single face, integrate over the spinors, and are hence left with a functional that we just call $Z_f[\varphi,A]$. This functional can depend only on the $SU(2)$ connection on the edges, and the Lagrange multiplier $\varphi$ imposing the length matching condition, all other variables have been integrated out:
\begin{equation}
Z_f[A,\varphi]:=\smashoperator{\int_{\substack{z^A(0)=z^A(1)\\w^A(0)=w^A(1)}}}\mathcal{D}[z]\mathcal{D}[w]
\E^{\int_{\partial f}\di t\big(z^\dagger_A\frac{D}{\di t}z^A-w^\dagger_A\frac{\di}{\di t}w^A-\I\varphi(\|z\|^2-\|w\|^2)\big)}.\label{faceampl}
\end{equation}
If we now want to recover the path integral for the full discretised manifold,  we just take the product over all face amplitudes \eref{faceampl} and integrate over all remaining configuration variables:
\begin{equation}
Z_M=\int\prod_{f:\text{faces}}\mathcal{D}[\varphi_f]\Delta_{\mathrm{FP}}^\psi[\varphi_f]\delta\big(\psi[\varphi_f]\big)
\prod_{e:\text{edges}}\mathcal{D}[A_e]\Delta_{\mathrm{FP}}^\Psi[A_e]\delta\big(\Psi[A_e]\big)Z_f[\varphi_f,A].
\end{equation}

The Lagrangian \eref{step8} for the face action $S_f[z,w,\varphi,A]$ is quadratic in the spinors. This considerably simplifies the evaluation of the path integral $Z_f[\varphi,A]$. We only have to calculate an infinite product of Gaußian integrals. This eventually yields a trace over the underlying Hilbert space:
\begin{equation}
Z_f[A,\varphi]=\mathrm{Tr}_{\mathcal{H}\otimes\mathcal{H}}\Big[\mathrm{Pexp}\Big(-\int_{\partial f}\di t\big(A^i(t)\ou{\tau}{AB}{i}\hat{z}_A\hat{z}^\dagger_B+\I\varphi(t)({:}\|\hat{z}\|^2{:}-{:}\|\hat{w}\|^2{:})\big)\Big)\Big].
\end{equation} 
The trace goes over an orthonormal basis in the Hilbert space $\mathcal{H}\otimes\mathcal{H}\ni f(z,w)$ of analytic functions in the $z$- and $w$-spinors, square integrable with respect to the inner product \eref{innprod}. In terms of the spin $(j,m)$-basis \eref{canbasis} this trace turns into an infinite sum:
\begin{equation}
Z_f[A,\varphi]=\sum_{2j=0}^\infty\sum_{m=-j}^j\sum_{2l=0}^\infty(2l+1)\big \langle j,m\big|\mathrm{Pexp}\Big(\I\int_{\partial f}\di t A^i(t)L_i\Big)\big|j,m\big\rangle\E^{-\I\int_{\partial f}\di t\varphi(t)(2j-2l)}.
\end{equation}

\emph{Step 2: Integration over the $U(1)$ gauge potential.} The next step is to perform the integrals over the gauge potentials. Let us first do the integral over $\varphi$. This requires a gauge fixing, and we choose the following:
\begin{equation}
\psi[\varphi](t)=\frac{\di}{\di t}\varphi(t)=0.\label{U1gaugefix}
\end{equation} 
The variation $\delta_\lambda=\frac{\di}{\di\varepsilon}|_{\varepsilon=0}$ of the gauge fixing condition \eref{U1gaugefix} under an infinitesimal $U(1)$ gauge transformation $\varphi_{\varepsilon\lambda}=\varphi+\varepsilon\dot{\lambda}$ determines the Faddeev--Popov determinant $\Delta_{\mathrm{FP}}^\psi[\varphi]$ as the functional determinant of the following differential operator $\hat{m}$:
\begin{equation}
\hat{m}[\lambda]:=
\frac{\di}{\di\varepsilon}\Big|_{\varepsilon=0}\psi[\varphi_{\varepsilon\lambda}](t)=\frac{\di^2}{\di t^2}\lambda(t).
\end{equation}
The eigenvectors of $\hat{m}$ are clearly independent of $\varphi$, and so is the Faddeev--Popov determinant $\Delta_{\mathrm{FP}}^\psi[\varphi]$, which can therefore only affect the overall normalisation of the measure. The gauge potential $\varphi$ determines a $U(1)$ angle, periodic in $2\pi$. We require that the integration measure is normalised, which in turn implies $\Delta_{\mathrm{FP}}^\psi[\varphi]=(2\pi)^{-1}$ once we restrict the integral over just one period of $\varphi$. The resulting integral gives the Dirac distribution\footnote{The Dirac distribution evaluates any $f:SU(2)\rightarrow\C$ at the identity $\mathds{1}$: $\int_{SU(2)}d\mu_{\mathrm{Haar}}(U)f(U)\delta_{SU(2)}(U)=f(\mathds{1})$, where $d\mu_{\mathrm{Haar}}(U)$ is the normalised Haar measure on the group.} of the holonomy around the spinfoam face:
\begin{align}
\nonumber Z_f[A]&=\int\mathcal{D}[\varphi]\Delta_{\mathrm{FP}}^\psi[\varphi]\delta\big(\psi[\varphi]\big)Z_f[A,\varphi]=\\
\nonumber &=\frac{1}{2\pi}\sum_{2l=0}^\infty\sum_{2j=0}^\infty\sum_{m=-j}^j\int_0^{2\pi}\di\varphi\,\E^{-\varphi(2j-2l)}(2l+1)\big \langle j,m\big|\mathrm{Pexp}\Big(\I\int_{\partial f}\di t A^i(t)L_i\Big)\big|j,m\big\rangle=\\
&=\sum_{2j=0}^\infty\sum_{m=-j}^j(2j+1)\big\langle j,m\big|\mathrm{Pexp}\Big(\I\int_{\partial f}\di t A^i(t)L_i\Big)\big|j,m\big\rangle=\delta_{SU(2)}\Big(\mathrm{Pexp}\Big(-\int_{\partial f}\di tA^i(t)\tau_i\Big)\Big),\label{faceampl2}
\end{align}
where the last equality follows from the Peter--Weyl theorem. 

\emph{Step 3: Integral over the $SU(2)$ gauge potential on the edges.} We are now left to perform the integral over the $SU(2)$ connection. Our strategy is to solve this path integral on each edge separately. The calculation can be seen as a one dimensional analogue of what has been found in \cite{Bianchi:2009tj, ReinhardtHaar}. In fact, Bianchi's conjecture \cite{Bianchi:2009tj} of equivalence between the Ashtekar--Lewandowski measure \cite{LOSTtheorem, thiemann} on a fixed graph and the canonical measure in the moduli space of flat connections was one of the key motivating ideas behind this work.

Simplifying our notation, let us first parametrise each edge $e, e^\prime,\dots$ by a $t$-coordinate running from $0$ to $1$. Every edge carries its own $SU(2)$ gauge potential $A_e(t)=A_\mu(e(t))\dot{e}^\mu(t)$, defined as in \eref{edgeconn}.
We now choose our gauge condition, and require that for every edge $e$ the gauge potential be constant in $t$:
\begin{equation}
\forall e: \Psi^i[A_e](t)=\frac{\di}{\di t}A^i_e(t)=0.\label{gaugecond}
\end{equation}
Notice that this is only a partial gauge fixing\footnote{A complete gauge fixing condition is inconvenient, because it would depend on the topological details of how the edges bound another. The gauge fixing \eref{gaugecond}, on the other hand, is more general. We can simultaneously impose it on every single edge, no matter how the edges glue together. The proof follows in a minute.}.  For a single edge $e$, residual gauge transformations can shift the connection $A_e\equiv A$ to any other constant value $\tilde{A}$. The proof is immediate, consider the gauge element:
\begin{equation}
g(t)=\E^{-At}g_o\E^{\tilde{A}t},\quad g_o\in SU(2),\quad\text{and}\quad A,\tilde{A}\in \mathfrak{su}(2).\label{resid}
\end{equation}
The gauge transformed connection yields the $\mathfrak{su}(2)$ element $\tilde{A}$, which is again constant in $t$:
\begin{equation}
\tilde{A}=g^{-1}(t)\dot{g}(t)+g^{-1}(t)Ag(t).
\end{equation}
A typical gauge invariant observable is the Wilson loop---the trace of the holonomy around the boundary of the spinfoam face. If the gauge condition \eref{gaugecond} holds on all edges, we have, in fact:
\begin{equation}
\mathrm{Tr}\Big(\mathrm{Pexp}\big(-\int_{\partial f}A\big)\Big)=\mathrm{Tr}\Big(\mathop{\raisebox{-2pt}{\text{\Large P}}\!\prod}_{e\in\partial f}\E^{-A_e}\Big),
\end{equation}
where ${\raisebox{-1.5pt}{\text{\large P}}\!\prod}$ denotes the  path ordered product\footnote{Let $\partial f$ consist of edges $e_i:[0,1]\rightarrow\partial M$. Their orientation agree with the induced orientation of $\partial f$, and they also be already appropriately ordered: $\forall i:e_{i+1}(0)=e_i(1)$, $e_{N+1}\equiv e_1$. We can now define the path ordered product simply as $\raisebox{-0.5pt}{}\mathrm{P}\!\prod_{i=1}^NU_{e_i}:=U_{e_N}U_{e_{N-1}}\cdots U_{e_1}$, where each edge carries the holonomy $\E^{-A_{e_i}}=U_{e_i}\in SU(2)$.\label{prodord}}.

We can impose the gauge fixing condition \eref{gaugecond} globally, on each individual edge of the discretisation. This can be seen as follows. Start with some generic gauge potential, not subject to \eref{gaugecond}. We now need a gauge transformation $g(t)$ mapping $A^i(t)$ into an element $\tilde{A}^i$ of the constraint hypersurface: $\Psi^i[\tilde{A}](t)=\frac{\di}{\di t}\tilde{A}^i(t)=0$. We compute the parallel transport $U(t)=\mathrm{Pexp}(-\int_0^t\di tA(t))$ along the edge, and define the $SU(2)$ angle $\phi^i$ as the logarithm of the holonomy along the entire edge: $U(1)=\exp(-\phi^i\tau_i)$. The gauge transformation
\begin{equation}
g(t)=U(t)\E^{t\phi^i\tau_i}\label{gaugetrafo}
\end{equation}
fulfils our requirement, for it turns the connection into a constant $\mathfrak{su}(2)$ element:
\begin{equation}
\tilde{A}(t)=g^{-1}(t)\dot{g}(t)+g^{-1}(t)A(t)g(t)=\phi^i\tau_i.
\end{equation}
This enables us to solve $\Psi^i[A]=0$ all along the edge. In fact, we can achieve \eref{gaugecond} on all edges $e, e^\prime,\dots$ at the same time, simply because the gauge transformation \eref{gaugetrafo} vanishes at the edge's source and target points: $g(0)=g(1)=\mathds{1}$.

The residual gauge transformations \eref{resid} preserve the gauge fixing condition $\Psi^i[A]=0$. The Faddeev--Popov procedure does not remove these \qq{horizontal} transformations, it only deals with transversal gauge transformations that take us out of the gauge fixing surface. Transversal gauge transformations vanish at the two boundary points $t=0,1$ of the edge $e(t)$, but not in between. We can now spilt any gauge element $\Lambda^i(t):[0,1]\mapsto \mathfrak{su}(2)$ into its horizontal and transversal components: $\Lambda^i(t)=\Lambda^i_\parallel(t)+\Lambda^i_\perp(t)$, where $\Lambda^i_\parallel(t)$ maps the gauge fixing surface into itself \eref{resid}, while $\Lambda_\perp^i(t)$ deforms it non-trivially.

Since any transversal gauge element $\Lambda^i(t)_\perp\equiv\Lambda^i(t)$ vanishes at the boundary, it implicitly defines a periodic function $\Lambda^i(t+n)=\Lambda^i(t)$ on the real line. We can thus introduce Fourier modes $\E^{2\pi\I nt}$ and write: 
\begin{equation}
\Lambda^i(t)=\sum_{n=1}^\infty\Lambda^i_n\E^{2\pi \I nt}+\CC,\quad \Lambda^i(0)=0,\quad \Lambda^i_n\in\C^3,\label{trans}
\end{equation}
where $\CC$ denotes the complex conjugate of all preceding terms. Notice the absence of the $n=0$ mode, which would generate residual gauge transformations preserving the gauge fixing surface $\Psi^i[A]=0$ \eref{gaugecond}. We exclude this constant gauge element because only the transversal modes, that map the gauge fixed connection out of the constraint hypersurface $\Psi^i[A]=0$, can contribute to the Faddeev--Popov determinant.

We now need the Faddeev Popov operator $\hat{M}$. Infinitesimal gauge transformations  \eref{gaugetrafo2} of the gauge fixing condition \eref{gaugecond} define this operator as  
\begin{equation}
\ou{\hat{M}}{i}{j}\Lambda^j(t)=\frac{\di}{\di \varepsilon}\Big|_{\varepsilon=0}
\big(\rho_{\varepsilon\Lambda}A\big)^i(t)=\frac{\di^2}{\di t^2}\Lambda^i(t)+\ou{\epsilon}{i}{jk}A^j\frac{\di}{\di t}\Lambda^k(t).
\end{equation}
Its eigenvalues determine the Faddeev--Popov determinant in the space of transversal gauge elements \eref{trans}.
Let us do the calculation for only one direction of $A^i$. Setting, without loss of generality, $A^i=A\delta^i_3$, we are thus led to the following eigenvalue equation:
\begin{equation}
\left(\begin{split}
\frac{\di^2}{\di t^2}\Lambda^1-A\frac{\di}{\di t}\Lambda^2&=E\Lambda^1\\
\frac{\di^2}{\di t^2}\Lambda^2+A\frac{\di}{\di t}\Lambda^1&=E\Lambda^2\\
\frac{\di^2}{\di t^2}\Lambda^3&=E\Lambda^3
\end{split}\right)
\end{equation}
The eigenvectors are $\Lambda^{(n)}_\pm$ and $\Lambda^{(n)}_z$ with corresponding eigenvalues $E^{(n)}_\pm$ and $E^{(n)}_z$ respectively:
\begin{align}
\Lambda^{(n)}_\pm(t)&=\frac{1}{\sqrt{2}}\Bigl(\begin{smallmatrix}1\\\pm\I\\0\end{smallmatrix}\Bigr)\E^{2\pi\I nt},&&\hspace{-1.5cm}\text{with:}\quad E^{(n)}_\pm=-(2\pi n)^2\Big(1\mp\frac{A}{2\pi n}\Big),\quad n\in \Z-\{0\},\\
\Lambda^{(n)}_z(t)&=\Bigl(\begin{smallmatrix}0\\0\\1\end{smallmatrix}\Bigr)\E^{2\pi\I nt},&&\hspace{-1.5cm}\text{with:}\quad E^{(n)}_z=-(2\pi n)^2,\quad n\in\Z-\{0\}.
\end{align}
Only transversal gauge transformations can contribute to the Faddeev--Popov determinant, and therefore the $n=0$ modes do not appear in this list. The functional determinant of $\hat{M}$ is the product over all eigenvalues. This badly diverges, but we can easily remove this infinity by looking at a regularised expression:
\begin{equation}
\Delta_{\mathrm{FP}}^\Psi[A]:=\frac{\det\hat{M}\big|_{A\phantom{=0}}}{\det\hat{M}\big|_{A=0}}=\prod_{n\in\Z-\{0\}}\Big(1-\frac{|A|}{2\pi n}\Big)^2=\frac{4\sin^2\big(\frac{|A|}{2}\big)}{|A|^2}.
\end{equation}
The square sinc function combines with the flat integration measure $d^3A$ to the Haar measure of $SU(2)$. Taking the parametrisation $U=\exp(-A^i\tau_i)$ of $U\in SU(2)$ and setting $|A|=\sqrt{\delta_{ij}A^i A^j}$ we find, in fact: 
\begin{equation}
d^3A\frac{4\sin^2\big(\frac{|A|}{2}\big)}{|A|^2}=\frac{4}{3}\mathrm{Tr}\big(U^{-1}\di U\wedge U^{-1}\di U\wedge U^{-1}\di U\big)\Big|_{U=\exp(-A^i\tau_i)}=32\pi^2\,d\mu_{\text{Haar}}(U).
\end{equation}
Our gauge condition \eref{gaugecond} has thus turned the functional integral into an ordinary integral over the group. Absorbing the overall normalisation $32\pi^2$ into the definition of the flat integration measure $\mathcal{D}[A]\propto\prod_{t\in[0,1]}d^3A (t)$ we get the following rule:
\begin{equation}
\int\mathcal{D}[A]\Delta_{\mathrm{FP}}^\Psi[A]\delta\big(\Psi[A]\big) f\big(\mathrm{Pexp}(-\smallint_eA)\big)=\int_{SU(2)}d\mu_{\text{Haar}}(U)\,f(U),\label{intform}
\end{equation}
where $f:SU(2)\rightarrow \C$ denotes some integrable function on the group.

We arrive at our final result once we repeat the calculation for all edges in the discretisation. Combining the integration formula \eref{intform} with our final expression \eref{faceampl2} for the face amplitude $Z_f[A]$, we see that the path integral $Z_M$ over the spinorial action \eref{pathint} eventually assumes a very neat form:
\begin{equation}
Z_M=\prod_{e:\text{edges}}\int_{SU(2)}d\mu_{\mathrm{Haar}}(U_e)\prod_{f:\text{faces}}\delta_{SU(2)}\Big(\mathop{\raisebox{-2pt}{\text{\Large P}}\!\prod}_{e\in\partial f}U_e\Big),\label{finresult}
\end{equation}
where ${\raisebox{-1.5pt}{\text{\large P}}\!\prod}$ again denotes the path ordered product (see footnote \ref{prodord}). Equation \eref{finresult} reproduces the Ponzano--Regge model for the simplicial manifold $M$. This is our final result. It proves the equivalence between our one-dimensional spinorial path integral \eref{pathint} and the discrete spinfoam approach \cite{ponzanoregge, Barrett:2008wh, FreidelReggeI}.

\section{Conclusion}
\noindent 
\emph{Summary.}  This article developed two results: First of all, we wrote the discretised Palatini action as a one-dimensional line integral. We then took that action and used it to define the path integral. The resulting amplitudes reproduced the Ponzano--Regge model.

Section \ref{newactn} gave the first result. To discretise the first-order action, we introduced a simplicial decomposition of the three-dimensional manifold. The equations of motion for a discretised field theory normally give a tangled system of difference equations. This would make it hard to speak about the symplectic structure, the Hamiltonian, the time evolution and the constraint equations of the theory---all of which are crucial elements for the quantisation program.\footnote{Despite this difficulty, Hoehn and Dittrich \cite{Dittrich:2013jaa,Hoehn:2014fka,Hoehn:2014wwa} have recently achieved impressive progress towards this goal.} Our partial continuum limit circumvents this difficulties. On each spinfoam face we performed a continuum limit in the $t$-variable parametrising the boundary of the spinfoam face. The resulting action \eref{step9} is a one-dimensional line integral over the one-skeleton of the underlying simplicial manifold. The action is a functional of the loop gravity spinors \cite{twist}, but depends also on connection variables. There is a $SU(2)$ connection on each edge, and a $U(1)$ connection $\varphi$ for each spinfoam face. All fields are continuous, but have support only on the one-dimensional edges of the simplicial complex. The action variation gave us the equations of motion, and we proved agreement with the continuum theory: The Gauß law \eref{gausslaw} is the discrete analogue of the torsionless equation \eref{torscons}, while the evolution equations \eref{evolveq} imply that the loop holonomy transports the spinors into themselves \eref{flatconst}. This represents the flatness constraint \eref{flatcons} in the discrete theory. We closed the classical part with two more comments: First of all, the equations of motion \eref{evolveq}, admit a Hamiltonian formulation \eref{hameq}. Then we also discussed the local gauge symmetries of the theory. The action has a diffeomorphism symmetry, since it does not depend on the actual parametrisation of the edges, but there is also a $U(1)$ symmetry for each face, and an internal $SU(2)$ gauge symmetry.

This was our first complex of results. The second part considered the quantisation of the theory, as developed in section \ref{pathintsec}. We started with a short review showing how to recover the loop gravity Hilbert space from the spinorial representation \cite{Livinerep}. The remaining part developed the path integral for the simplicial manifold. 
The integral over the spinors is easy to solve: The action \eref{step8} is quadratic in the spinors, and the path integral reduces to an infinte product of Gaußian integrals. Then there are the local gauge symmetries. We removed the redundant integrals according to the usual Faddeev--Popov procedure. The result reduced the functional integral to an ordianry integral, with the emergence of the canonical Haar measure of $SU(2)$. Our final expression \eref{finresult} agrees with the Ponzano--Regge state sum model \cite{ponzanoregge, Barrett:2008wh,FreidelReggeI,FreidelReggeII,FreidelReggeIII} of three-dimensional quantum gravity.

\emph{Prospects of the formalism.} Our analysis encourages further questions: What is the physical interpretation of the $U(1)$ winding number $n_f$ as defined in \eref{windnumb}? Does the edge Hamiltonian \eref{edgeham} introduce a local notion of energy? What is the spinorial representation of the\emph{shift symmetry} $e^i\rightarrow e^i+D\lambda^i$, and can we use it to remove the divergencies of the Ponzano--Regge model through the usual Faddeev--Popov procedure \cite{FreidelReggeI}? 
Is there a way to add a cosmological constant to the spinorial action (along the lines of e.g. \cite{Noui:2011im,Pranzetti:2014xva,Dupuis:2013haa})? Can we generalise the formalism to the Lorentzian signature, thus replacing $SU(2)$ by $SU(1,1)$?  Can we use the one-dimensional action to bring causal sets \cite{Sorkin:2003bx} and spinfoams closer together, an idea first studied in \cite{Cortes:2013pba, fotini}?

The most important question is however rather simple: What does all this machinery actually tell us for the four-dimensional Lorentzian case? So far, we only have a partial answer: In 3+1 dimensions there is a spinorial (or rather twistorial) formulation of $SL(2,\C)$ $BF$-theory \cite{hamspinfoam}. Once again the action is a \emph{one-dimensional} integral over the edges of the discretisation. This action defines a topological theory. Additional constraints break the topological symmetries and bring us back to general relativity \cite{selfdualtwo, LQGvertexfinite}. How these simplicity constraints translate into conditions on the spinors has been shown in  \cite{komplexspinors,twistintegrals}. These results can lead us to a version of first-order Regge calculus in four spacetime dimensions with spinors as the fundamental configuration variables. Generalising our derivation of the Ponzano--Regge model to the four-dimensional case would then give us a neat definition of the transition amplitudes: We would start with the one-dimensional spinorial path integral for $SL(2,\C)$ $BF$-theory, add the simplicity constraints, and evaluate the integral for a given simplicial manifold. My hope is that this will improve our understanding of the mathematical structure of loop quantum gravity, its causal structure, and the continuum limit of the theory. 


\emph{Acknowledgements.}  I thank Eugenio Bianchi and Marc Geiller for helpful comments and discussions on a first draft of this article. I also gratefully acknowledge support from the Institute for Gravity and the Cosmos, the National Science Foundation through grant PHY-12-05388 and the Eberly research funds of
The Pennsylvania State University.



\appendix
\providecommand{\href}[2]{#2}\begingroup\raggedright\endgroup
\end{document}